# The Proof-Theoretic Origin of Double Negation Introduction & Elimination

**Khashayar Irani**
**Birkbeck College, University of London**
k.irani@mathematicallogic.com
September 2025

**Abstract**

This paper investigates the proof-theoretic foundations of double negation introduction (DNI: A ⊢ ¬¬A) and double negation elimination (DNE: ¬¬A ⊢ A) in classical logic.[1] By examining both sequent calculus and natural deduction, it is shown that these rules originate in reductio ad absurdum (RAA): DNI results from deriving ¬¬A via ¬I by discharging [¬A], while DNE arises from deriving A through a reductio on [¬A].[2] Their significance extends beyond semantic equivalence, for DNI and DNE embody the identity relation A ↔ ¬¬A as a structural principle of classical logic. The paper demonstrates that both rules possess harmony, ensuring balance between introduction and elimination, and normalisation, which guarantees that derivations reduce to canonical form without detours. These features reveal double negation not as a redundancy, but as a mechanism of proof-theoretic stability, securing the disciplined integration of RAA into classical logic.

*Keywords: Classical Logic, Double Negation, DNI, DNE, Reductio ad Absurdum, Sequent Calculus, Natural Deduction,*

## Introduction

In Section 1, we argue that the identity relation A ↔ ¬¬A is a foundational principle of classical logic, not a peripheral equivalence. This identity secures both the semantic principle of bivalence and the proof-theoretic legitimacy of RAA. To affirm A ↔ ¬¬A is to recognise that every proposition is interchangeable with the denial of its denial, ensuring that contradictions can be discharged into positive assertions.[3] We argue that this relation functions as the bridge between direct and indirect reasoning, underpinning the legitimacy of DNI and DNE. Without the stability provided by this identity, RAA would collapse into a purely negative tool, incapable of generating assertions. Thus, Section 1 establishes the indispensability of the identity $A \leftrightarrow \neg\neg A$, showing that it grounds both the semantic and inferential structure of the classical system. In Section 2, we argue that DNI and DNE themselves originate in the inferential machinery of RAA and that they can be formally derived within both sequent calculus and natural deduction. Beginning with the reflexive sequent $A \Rightarrow A$, DNI is proved by applying L¬ to yield $A, \neg A \Rightarrow$, deriving contradiction, and then applying R¬ to conclude $A \Rightarrow \neg\neg A$. In natural deduction, this is mirrored by assuming [A] and [¬A], deriving ⊥ via ¬E, and discharging [¬A] through ¬I to obtain ¬¬A. Conversely, we argue that DNE is derived from $A \Rightarrow A$ by applying R¬ to derive ¬A, which, when combined with [¬¬A], produces contradiction and allows the discharge of ¬A to restore A. The corresponding natural deduction structure replicates this eliminative process, yielding ¬¬E. We further argue that these rules embody the structural virtues of harmony and normalisation. Harmony guarantees reciprocity: what ¬¬I introduces, ¬¬E eliminates. Normalisation ensures that detours through ¬¬A collapse to direct proofs of A, demonstrating procedural economy. Together, these features show that reductio operates not as an unwieldy appendage but as a disciplined, balanced mechanism fully integrated into classical logic. Finally, in Section 3, we argue that these findings converge to reaffirm the indispensability of $A \leftrightarrow \neg\neg A$. Semantically, this identity secures bivalence, ensuring that the truth-value of A always coincides with that of ¬¬A. Proof-theoretically, it legitimises RAA by enabling contradiction to be transformed into positive assertion. We conclude that DNI and DNE are not redundant curiosities but structural principles essential to the consistency, completeness, and stability of classical derivability.

---

[1] To keep this paper as concise as possible, and in order to focus solely on formulating our thesis regarding DNI and DNE, we have refrained from engaging with similar views on these two classical principles. However, we fully acknowledge that such views exist, and that some may stand in conflict with our own interpretation. A more comprehensive comparison between these alternative perspectives and our account may be undertaken in future work.

[2] As usual, formulae placed within [ ] are considered assumed formulae, in contrast to derived or inferred formulae, which are presented without the use of [ ].

## 1 The Classical Foundations of The Identity of Double Negation

---

[3] Since this work has been chiefly composed for a proof-theoretic audience, some philosophical concepts such as "assertion" and "denial", which are usually treated as speech acts within the analytic tradition of the philosophy of language, may be employed here in a somewhat loose manner. By "assertion" we mean a positive formula, such as A, which does not display a negation; conversely, by "denial" we mean a negated formula, for example ¬A. Thus, at least on the surface, our use of the terms "assertion" and "denial" does not imply a bilateral understanding of logic, unless we explicitly state otherwise.

The identity relation A ↔ ¬¬A occupies a uniquely privileged position in classical logic. It does not merely stand as one among many equivalences derivable within the system; rather, it encapsulates a structural insight that defines the essence of classical derivability itself. To affirm that A and ¬¬A are interchangeable is to assert that a proposition is logically indistinguishable from the denial of its denial. This is not simply a matter of definitional convenience but a principle that secures the entire architecture of semantic interpretation, inference, and proof within the classical framework. It demarcates the boundary between classical and non-classical traditions, particularly intuitionistic logic, where the equivalence is not accepted in full generality. However, to appreciate the centrality of A ↔ ¬¬A, we must first recall the mechanics of negation itself in proof theory.[4]

In proof-theoretic terms, negation is typically governed by introduction and elimination rules namely, ¬I:

$$\begin{array}{c} [A] \\ \Pi \\ \bot \\ \hline \neg A \end{array} \neg I$$

by which a contradiction derived from assuming [A[ licenses ¬A; and ¬E:

$$\begin{array}{cc} A & [\neg A] \\ \hline \bot \end{array} \neg E$$

by which the joint presence of A and [¬A] yields contradiction (⊥).[5] Within this structural context, the equivalence A ↔ ¬¬A arises from the repeated application of these rules in a manner that renders them mutually reinforcing. When the rules are considered in isolation, one may question whether affirming A truly commits us to the rejection of ¬¬A; yet when embedded within a proof system that validates RAA, the equivalence emerges as inevitable. The reductio principle is key here since its allows us to derive A by assuming its negation ([¬A]) and deriving a contradiction.[6] This inferential move depends on precisely the symmetry expressed by the identity relation A ↔ ¬¬A. If denying A entails contradiction, then affirming A follows immediately, since to sustain the negation of A would itself negate the consistency of the proof. Conversely, to affirm ¬¬A is simply to affirm that denying A cannot be done without collapsing into contradiction. Thus, the proof-theoretic power of RAA presupposes the identity relation between a formula and its double negation.

---

[4] For a detailed and technical survey on negation, see Dov M. Gabbay & Heinrich Wansing (1999) and Viviane Déprez & M. Teresa Espinal (2020).

[5] This natural deduction presentation of ¬E, in which only the bracketed formula [¬A] is introduced as an assumption and its negated counterpart, namely A, is shown as a derived formula, has been extracted from a context-dependent sequent calculus system. In this system, A is derived from the context Γ in the form Γ ⇒ A, while [¬A] is introduced as an assumption by applying the L¬ rule to A, thereby obtaining the resulting sequent ¬A, Γ ⇒ ⊥. By contrast, if we were employing a context-free sequent calculus formalism and had applied the L¬ rule to the initial sequent A ⇒ A, thereby attaining the sequent A,¬A ⇒, then both A and ¬A would have to be represented as assumptions, that is, as [A] and [¬A]. Furthermore, since the stated ¬E rule corresponds to the displayed ¬I rule, A must be presented as a derived formula and ¬A as an assumed one ([¬A]).

[6] In this example, the sequent calculus form of our RAA proof is ¬A,Γ ⇒ ⊥ ⊢ Γ ⇒ A, where we first assume [¬A] and, since this assumption leads to a contradiction, we therefore derive its negated counterpart, namely A.

From the semantic perspective, the equivalence A ↔ ¬¬A directly expresses the principle of bivalence, which asserts that every proposition is either true or false. Under bivalence, the truth-value of ¬¬A must coincide with that of A: if A is true, then ¬A is false, making ¬¬A true; if A is false, then ¬A is true, making ¬¬A false. In either case, no gap arises between the evaluation of A and ¬¬A. Classical semantics thus ensures that truth-values remain stable under double negation, preventing the sort of divergence characteristic of intuitionistic models, where ¬¬A may hold without entailing A. The identity relation thus embodies the deep interconnection between proof theory and semantics in the classical tradition. Likewise, the proof-theoretic import of the equivalence cannot be overstated. By treating A and ¬¬A as interderivable, a classical system allows proofs to transition seamlessly between direct and indirect forms of reasoning. A proof that establishes ¬¬A through contradiction is effectively equivalent to a proof of A itself. This symmetry ensures that indirect proofs, particularly those employing RAA, can be regarded as fully legitimate pathways to positive conclusions. Without A ↔ ¬¬A, the transition from contradiction back to affirmation would lack justification, and indirect proof methods would be deprived of their foundational legitimacy.

Indeed, the very possibility of DNE, the inference from ¬¬A ⊢ A, depends upon this identity because it can be viewed as the operational expression of the semantic equivalence between A and ¬¬A.[7] It affirms that the indirect route of establishing ¬¬A does not merely suspend the denial of A, but conclusively establishes A itself. Without DNE, classical proofs would fragment into direct and indirect domains with no guarantee of convergence. The capacity to eliminate double negation, therefore, underwrites the consistency of the entire inferential network. At the same time, DNI that is, the inference from A ⊢ ¬¬A, equally exemplifies the identity. This rule validates the intuition that affirming A suffices to deny the possibility of denying it. It ensures that once a proposition is established, its double negation can be introduced without further proof. DNI thus mirrors DNE, creating a bidirectional structure that secures the equivalence A ↔ ¬¬A. The harmony of these two rules guarantees that logical consequence is preserved under both expansion and contraction of negations, ensuring

---

However, in this work, from both a sequent calculus and a natural deduction perspective, the form of RAA employed in the subsequent proofs of the next section (as well as the above example) is A,Γ ⇒ ⊥ ⊢ Γ ⇒ ¬A. In sequent calculus, this corresponds to the rule R¬, while in natural deduction it matches the ¬I principle.

[7] Although the title of this paper concerns the proof-theoretic analysis of the rules of double negation introduction and elimination, and while in a natural deduction framework it is customary first to discuss the introduction rules associated with a constant and only thereafter its elimination rules, in this section we shall proceed in the reverse order by considering DNE before DNI. The reason for this is that, as will become evident in the next section, DNE serving as a shorthand for the rules formulated in sequent calculus and natural deduction as R¬ and ¬I, is proof-theoretically derived prior to DNI, which is formally grounded on L¬ and ¬E.

stability of proof transformations.[8] The indispensability of the equivalence becomes evident when we consider what collapses in its absence. Suppose we were to deny A ↔ ¬¬A. Then proofs of ¬¬A could not be transformed into proofs of A, and RAA would lose its authority as a general inferential rule. Similarly, DNI would no longer preserve the closure of assertion under negation. The absence of the equivalence would therefore dismantle the connection between contradiction, denial, and assertion, leaving the system incomplete. Many canonical results of classical logic, from the law of the excluded middle (LEM) to the general form of reductio, depend upon this structural identity. Consequently, it follows that A ↔ ¬¬A must be regarded as more than a contingent feature of the classical system. It is a cornerstone of its logical architecture, grounding both the semantic principle of bivalence and the proof-theoretic principle of RAA. The identity ensures that proofs are not only consistent but also complete, in the sense that any valid inference can be reconstructed through the interplay of negation and double negation. It thus functions as a bridge between the two poles of classical logic: the semantic guarantee that every proposition is determinately true or false, and the proof-theoretic assurance that contradiction can always be discharged in favour of assertion or denial. Nonetheless, in order to understand how this cornerstone operates in detail, it is instructive to examine its dual manifestations: first, as a semantic equivalence grounded in truth-value assignments; and second, as a proof-theoretic mechanism realised through the application of inference rules. Both aspects converge to establish the indispensability of A ↔ ¬¬A. On the semantic side, we encounter the absoluteness of bivalence and the impossibility of divergence between A and ¬¬A. On the proof-theoretic side, we see how the rules of sequent calculus and natural deduction systematically generate and rely upon this equivalence. To take the semantic perspective first, let us consider a standard valuation in classical propositional calculus. For any proposition, we define its value as either true or false. The negation of a proposition is true if and only if the original proposition is false. By repeating this process, the double negation of a proposition is true if and only if the negation of the proposition is false. But the negation of a proposition is false if and only if the original proposition is true. Therefore, the double negation of a proposition is true if and only if the original proposition is true. Similarly, the double negation of a proposition is false if and only if the original proposition is false. Hence, for every valuation, the truth-value of a proposition always coincides with the truth-value of its double negation. This semantic identity directly entails that the equivalence between a proposition and its double negation is valid across all models of classical logic. The equivalence is therefore not a contingent or derived fact but a semantic theorem that expresses the determinacy of classical truth. On the other hand, from the proof-theoretic side, the equivalence is secured by the harmony of DNI and DNE. As we shall see in the next section, to prove DNI, we assume [A], derive contradiction from assuming [¬A], and then discharge [¬A] to infer ¬¬A. Conversely, to prove DNE, we assume [¬¬A], use RAA to derive contradiction from assuming [A], and thereby discharge the assumption to infer ¬A. In both cases, the mechanism of negation introduction operates to link assertion and denial through the possibility of contradiction. What differs is only the direction of the inference: DNI secures the passage from assertion to double denial, while DNE secures the passage from double denial to assertion. Together, they constitute the inferential realisation of the semantic identity. Thus, the identity relation A ↔ ¬¬A is indispensable for both semantic lucidity and proof-theoretic functionality in classical logic. It validates the equivalence of direct and indirect proofs, legitimises RAA as a fundamental rule of inference, and guarantees closure under negation. Without it, the system would lack both consistency and completeness. With it, the system achieves a structural harmony that unifies contradiction, denial, and assertion into a sound whole.

---

[8] In the next section, we shall demonstrate that DNI and DNE are well-behaved rules, as they constitute harmonious and normalisable classical principles.

Furthermore, the indispensability of the equivalence between a proposition and its double negation also comes into sharper focus when we turn to the history of proof theory and its formalisation in the twentieth century. Gerhard Gentzen's (1935) natural deduction and sequent calculus systems both rely implicitly on the symmetry between A and ¬¬A. While Gentzen did not present A ↔ ¬¬A as a primitive axiom, the derivations he permitted through rules of negation introduction and elimination presuppose that double negation functions as a bridge between contradiction and assertion. In fact, Gentzen's cut-elimination theorem, one of the cornerstones of proof theory, depends on the stability of derivations under the presence or absence of double negation. If derivations could not be normalised in this manner, then the structure of proof reduction itself would lose lucidity. This observation suggests that A ↔ ¬¬A should not be seen merely as a theorem internal to classical logic, but as a reflection of the meta-theoretical properties of proof systems. The ability to introduce and eliminate double negation without loss of information ensures that the calculus remains stable under structural transformations. In practice, this stability allows us to treat indirect arguments as equivalent in strength to direct ones, thereby broadening the scope of classical logic while ensuring that derivations do not proliferate uncontrollably. Without this property, many proofs would become stranded in a kind of inferential limbo, unable to progress from contradiction back to assertion. Moreover, another illuminating way of appreciating the importance of A ↔ ¬¬A is to contrast the situation in intuitionistic logic. In the intuitionistic tradition, DNI (A ⊢ ¬¬A) is provable, but DNE (¬¬A ⊢ A) is not, except in restricted cases. This asymmetry reflects the rejection of LEM and the broader constructivist orientation of intuitionistic logic. Here, the denial of a denial is not taken as sufficient to affirm a proposition, since such an affirmation requires a constructive proof of A itself. The contrast reveals that the acceptance of ¬¬A ⊢ A is precisely what gives classical logic its distinctive strength that is, it allows contradiction to be discharged in favour of affirmation, rather than leaving the matter unresolved. Therefore, the equivalence A ↔ ¬¬A can be seen as the logical fingerprint of classical logic.

The connection between the equivalence and RAA further reinforces its classical character. The RAA principle functions by showing that the denial of a proposition leads to absurdity, and hence that the proposition must hold. But unless ¬¬A collapses into A, such derivation would not be valid. Intuitionistic logic permits the weaker conclusion that denying A is impossible, but does not permit the stronger conclusion that A must therefore be true. Classical logic, by contrast, validates the stronger claim because it assumes that for every proposition A, either A or ¬A must hold. Thus, the equivalence A ↔ ¬¬A can be regarded as the proof-theoretic codification of RAA. Without it, reductio would be restricted to a purely negative function; with it, reductio becomes a positive engine of proof, capable of generating assertions from contradictions. However, from another angle, the equivalence also underwrites the relationship between consistency and completeness in classical systems. Consistency requires that contradictions not be derivable, while completeness requires that every valid inference be derivable. The identity relation A ↔ ¬¬A ensures that no gap arises between the semantic fact that truth-values are bivalent and the proof-theoretic fact that proofs can recover assertions from contradictions. If DNE was not valid, classical logic would no longer be complete, since there would be truths of the form A that could not be recovered from the proofs of their double negations. Conversely, if DNI was not valid, consistency would be jeopardised, since assertions would not guarantee the denial of their denials. The two halves of the equivalence thus work together to secure both pillars of logical adequacy. Additionally, the equivalence of A and ¬¬A also impacts the understanding of structural rules such as weakening, contraction, and cut. Because A ↔ ¬¬A ensures that proofs are invariant under double negation, the structural rules can operate without threatening the stability of derivations. Weakening, for instance, allows the addition of irrelevant assumptions without affecting derivability. But if double negation were not equivalent, weakening could inadvertently introduce assumptions that undermine the system's closure under negation. Similarly, contraction relies on the interchangeability of multiple copies of the same formula, a property that presupposes that A and ¬¬A can substitute for one another without altering validity. The equivalence thus provides the background condition for the proper functioning of structural rules.

Eventually, as we prepare to move into a detailed analysis of DNI and DNE, it is worth pausing to summarise the significance of A ↔ ¬¬A. From the semantic perspective, it secures the principle of bivalence, ensuring that propositions and their double negations share the same truth-value in every model.[9] From the proof-theoretic perspective, it validates the legitimacy of RAA and ensures the stability of derivations under transformations involving negation. From the structural perspective, it provides the foundation for the proper operation of rules such as weakening, contraction, and cut. It is precisely for these reasons that the equivalence can be described as a cornerstone of classical logic. Its absence would deprive RAA of its power, fragment the inferential network, and compromise the system's consistency and completeness. Its presence, by contrast, ensures that classical logic is a robust and consistent framework, capable of supporting both direct and indirect proofs. The stage is therefore set for a closer examination of the two rules that operationalise this equivalence. In the next section, we shall show how these rules can be derived within a context-free sequent calculus and represented within natural deduction. In so doing, we will demonstrate that their proof-theoretic origin lies in the RAA structure, which itself finds its legitimacy in A ↔ ¬¬A.

---

[9] Philosophically speaking, it is also instructive to consider the implications of this equivalence for the concept of truth. In classical logic, truth is absolute and determinate: a proposition cannot hover between affirmation and denial. The equivalence A ↔ ¬¬A encodes this determinacy. To deny the denial of a proposition is, in effect, to assert the proposition. The two are indistinguishable. This is not merely a statement about the mechanics of proof but an affirmation of a philosophical stance: that truth and falsity are mutually exclusive and jointly exhaustive. By contrast, the rejection of DNE in intuitionistic logic reflects a different conception of truth, one that is constructive and open-ended. Thus, the identity of double negation can be seen as the hinge on which two entire philosophies of logic turn.

## 2 The Derivation of Double Negation Introduction & Elimination

In this section, we prove the claim that within both sequent calculus and natural deduction frameworks, DNI and DNE are respectively proved through applications of the R¬ and ¬I rules, each of which constitutes an instance of RAA. To substantiate our claim, consider the following two sequent calculus proofs of DNI and DNE:[10]

$$
\begin{array}{cc}
\dfrac{A \Rightarrow A}{\dfrac{A, \neg A \Rightarrow}{A \Rightarrow \neg\neg A} \, R\neg} \, L\neg
&
\dfrac{A \Rightarrow A}{\dfrac{\Rightarrow A, \neg A}{\neg\neg A \Rightarrow A} \, L\neg} \, R\neg
\end{array}
$$

As it is understandable from the above proof of DNI, after presenting the initial state, namely the reflexive sequent A ⇒ A, through an application of the rule L¬, we began by introducing the two assumed formulae [A] and [¬A] as our initial assumptions in the shape of the sequent A,¬A ⇒, and as a result of this empty-succedent sequent, derived a contradiction.[11] To display this formal step in a natural approach, consider the following natural deduction derivation:[12]

$$
\dfrac{[A] \quad [\neg A]}{\bot} \, \neg E
$$

in which, via an instance of the rule ¬E, we discharged the two assumptions [A] and [¬A] and derived a contradiction (⊥). Subsequently, in our sequent calculus formalism, we applied the rule R¬ on the already negated assumed formula [¬A] and derived the doubly negated formula ¬¬A as our objective formula. At the final stage of our proof, by attaining the endsequent A ⇒ ¬¬A, we managed to prove the rule DNI in a context-free sequent calculus setting. In the sequent calculus proof of DNI, the endsequent A ⇒ ¬¬A, is best understood as an extension of the identity relation A ↔ ¬¬A within the proof-theoretic framework. Starting from the basic identity A ⇒ A, the sequent A ⇒ ¬¬A demonstrates that identity is preserved even under the introduction of a double negation. In structural terms, the proof of DNI proceeds by taking the initial sequent A ⇒ A, introducing a contradictory state A,¬A ⇒, and then closing it through the application of the right negation rule to obtain A ⇒ ¬¬A. This shows that the reflexivity of identity is not confined to the simple repetition of A, but can be transformed into a statement about the stability of A under negation. The sequent A ⇒ ¬¬A thus represents a proof-theoretic reinforcement of identity, indicating that once A is given, the system can consistently sustain its double negation. In this way, the proof of DNI does not merely restate the triviality of identity, but reconfigures it through the reductio structure embedded in the rules of negation. The resulting sequent therefore expresses that the identity relation has a deeper structural robustness: not only does A stand in identity with itself, but the inferential machinery ensures that A also projects forward into the stability condition expressed by ¬¬A. It is at this point, after establishing DNI, that the relation to its converse and hence to the full equivalence A ↔ ¬¬A becomes significant. Subsequently, to show the natural progress of our sequent proof, in a natural deduction atmosphere, we can demonstrate the complete derivation of DNI via the following construction:

$$
\begin{array}{c}
[A] \quad\quad [\neg A] \\
\hline
\quad\quad\quad\quad\quad\quad \neg E \\
\bot \\
\hline
\quad\quad\quad\quad\quad\quad \neg I \\
\neg\neg A
\end{array}
$$

---

[10] Although it does not concern the aims of this paper, it is worth pointing out, for future reference, that the context-dependent forms of DNI and DNE in a singlesuccedent sequent calculus framework are, respectively, Γ ⇒ A ⊢ Γ ⇒ ¬¬A and Γ ⇒ ¬¬A ⊢ Γ ⇒ A.

[11] Within proof theory, it is recognised that a sequent such as A,¬A ⇒, which possesses an empty succedent, is equivalent to a sequent such as A,¬A ⇒ ⊥, in which the succedent is explicitly filled with ⊥.

[12] Keep in mind that this ¬E rule, which enables us to derive a contradiction from the two assumptions [A] and [¬A], is based on a context-free sequent calculus construction whose initial sequent is A ⇒ A, and not the sequent A,Γ ⇒ A. Therefore, as stated above, the sequent calculus form of this natural deduction derivation is A,¬A ⇒, and not Γ ⇒ A ⊢ ¬A,Γ ⇒ ⊥.

Prior to moving to the analysis of DNE, from the sequent calculus proof of DNI and the natural deduction formalism set out above, we can extract a compressed yet proof-theoretically valid natural deduction inference rule for DNI in the following shape:[13]

$$
\begin{array}{c}
A \\
\hline \neg\neg I \\
\neg\neg A
\end{array}
$$

To understand the formation of this rule, first consider DNI in the form of the endsequent A ⇒ ¬¬A, which, as explained above, is logically valid in virtue of the validity of the identity relation A ↔ ¬¬A. On the basis of this endsequent, the doubly negated ¬¬A is obtained directly from A. Accordingly, when examining our natural deduction proof of DNI, we must locate the part of the derivation that yielded A ⊢ ¬¬A. Since the derivation of ¬¬A required discharging the assumption [¬A] by an application of the rule ¬I, the conclusion, namely ¬¬A, must be understood as the doubly negated result of the remaining assumption [A]. Additionally, because the identity relation A ⇒ ¬¬A plays a fundamental role in the proof of DNI, in other words, because DNI demonstrates a doubly negated reflexive relation in the endsequent A ⇒ ¬¬A, we are proof-theoretically justified in transforming this reflexive endsequent into a natural deduction rule of doubly negated identity. Finally, since the discharging of the assumption [¬A] by ¬I results in the conclusion ¬¬A, the standard negation introduction rule ¬I is accordingly adapted into a rule of double negation introduction, expressed as ¬¬I.

---

[13] The reason we state "a compressed yet proof-theoretically valid natural deduction inference rule" is that, from this point onwards, we may justifiably employ our DNI inference rule A ⊢ ¬¬A instead of relying on the stated natural deduction proof whose final conclusion was the doubly negated formula ¬¬A. Of course, it should be borne in mind that whenever we employ this DNI inference rule, it presupposes, in the background, that we have already accepted the classical identity relation A ↔ ¬¬A together with the context-free sequent calculus proof of DNI. However, the well-behaved employment of DNI in deriving classical results may, perhaps, be deferred to future works.

On the other hand, in the sequent calculus derivation of DNE, after establishing the preliminary reflexive sequent A ⇒ A, we first derived the formula ¬A through an application of the rule R¬, which is the sequent calculus form of ¬I or, RAA. In a natural deduction framework, this RAA derivation can be represented as follows:

$$
\begin{array}{c}
[A] \\
\hline
\begin{array}{cc}
A \vee B & [\neg(A \vee B)]
\end{array} \quad \vee I \\
\hline
\bot \quad \neg E \\
\hline
\neg A \quad \neg I
\end{array}
$$

Next, following the derivation of ¬A in our sequent calculus proof, we introduced the assumption [¬¬A]. Since this assumption, when combined with the previously derived ¬A, yields a contradiction, a feature that cannot be directly displayed in a sequent calculus framework but can only be made explicit within a natural deduction setting, we discharged ¬A and thus derived A. This process, consisting of the assumption [¬¬A], the discharge of ¬A, and the derivation of A, can be properly represented within the natural deduction framework as follows:[14]

---

[14] Greg Restall, in both his unpublished paper *Speech Acts & the Quest for a Natural Account of Classical Proof* (2021) Page: 22, and his published article *Structural Rules in Natural Deduction with Alternatives* (2023) Page: 122, offers a natural deduction proof of DNE which, at first sight, appears structurally akin to the derivations we have presented. However, his proof proceeds by introducing store (↑) and retrieve (↓) operations on assumptions, and by marking assumptions such as [A̶] with syntactic strokes.

```
        [A]
    ─────────── ∨I
      A ∨ B         [¬(A ∨ B)]
    ──────────────────────────── ¬E
                 ⊥
    ──────────────────────────── ¬I
          ¬A            [¬¬A]
    ──────────────────────────── ¬E
                 ⊥
    ──────────────────────────── ¬I
                 A
```

The sequent calculus proof of DNE concludes with the endsequent ¬¬A ⇒ A, which functions as the eliminative counterpart to the constructive result of DNI. While DNI extended the reflexive sequent A ⇒ A into the form A ⇒ ¬¬A, thus proving that identity is stable under the introduction of a double negation, DNE demonstrates the reverse operation: that a doubly negated antecedent reduces back into the positive assertion of the original formula. In structural terms, DNE captures the proof-theoretic fact that the reflexive relation, once enveloped by negation, can always be restored to its original form. The proof itself reveals that the derived formula ¬A, once set against the assumption [¬¬A] in the antecedent, yields an immediate contradiction. This contradiction is then discharged through the application of the R¬ rule, resulting in the conclusion A. What is demonstrated here is that the derivation of ¬A cannot consistently stand alongside the assumption [¬¬A], and therefore the reflexive identity of A must be reinstated. Just as the endsequent A ⇒ ¬¬A in DNI was parasitic upon the reflexive axiom A ⇒ A, so too is the endsequent ¬¬A ⇒ A grounded in the same reflexive basis. Without the foundational stability of the identity axiom, the eliminative move that collapses a double negation back to A could not be justified. Consequently, DNE reveals that identity is not only preserved under the extension of negation but is also restored once the negation is removed.

From this perspective, the natural deduction structure of DNE can be abstracted into a compressed inference rule that directly reflects its eliminative nature. After deriving ¬A through an RAA subproof, the assumption [¬¬A] leads immediately to a contradiction, and this contradiction is discharged to yield A. The outcome can therefore be expressed in natural deduction as the following inference rule of double negation elimination:[15]

```
¬¬A
─────── ¬¬E
  A
```

---

[15] The same points explained in footnote 13 concerning DNI, also applies to DNE.

This rule mirrors the structure already exhibited in the derivation: the assumption of [¬¬A] suffices to conclude A, since the attempt to negate A is untenable in the presence of its double negation. In the same manner that we introduced ¬¬I as the natural deduction rule of double negation introduction, here we introduce ¬¬E as its eliminative counterpart. This rule is justified proof-theoretically by the endsequent ¬¬A ⇒ A, which captures the collapse of negation into identity. Taken together, the rules ¬¬I and ¬¬E:

```
   A                    ¬¬A
 _____ ¬¬I            _____ ¬¬E
  ¬¬A                   A
```

demonstrate that the identity relation A ↔ ¬¬A is not merely a formal coincidence but a structural principle within classical logic. Where ¬¬I showed that once A is established, its double negation may be freely introduced, ¬¬E now shows that the presence of a double negation suffices to eliminate it in favour of A. Thus, the constructive and eliminative halves converge to reinforce the reflexivity of identity, proving that A and ¬¬A are interderivable. In this manner, the mutual reinforcement of DNI and DNE elevates identity from a trivial axiom to a structural principle resilient under both the introduction and elimination of double negation.

To conclude this section, in what follows we will illustrate that DNI and DNE are harmonious and normalisable classical rules. At first sight, these structural features might appear to be little more than elegant ornaments of proof theory, valuable perhaps for aesthetic reasons but not indispensable to the functioning of the system. Yet this impression is misleading. When examined closely, harmony and normalisation reveal themselves to be indispensable to the inferential discipline of classical logic. They show that double negation is not a superfluous redundancy but a structural mechanism that secures both the internal balance and the procedural economy of proofs. To begin with, harmony ensures that the introduction and elimination of double negation align in such a way that neither outruns the content provided by the other. When DNI introduces ¬¬A from A, and DNE subsequently reduces ¬¬A back to A, the two rules form a perfectly reciprocal pair. The introduction rule does not commit us to more than can be eliminated, and the elimination rule does not demand resources that the introduction rule fails to provide. This equilibrium is not accidental: it reflects the underlying identity relation A ↔ ¬¬A. The harmony of these rules is therefore not a mere decorative symmetry but a structural necessity. It ensures that the rules of inference associated with negation do not destabilise the system but instead reinforce its consistency, permitting RAA to operate in a disciplined manner within classical reasoning.

A further dimension of harmony is its capacity to safeguard against inferential inflation. Without harmony, introduction and elimination rules can generate discrepancies that threaten to destabilise a proof system. For example, in the case of disjunction, poorly calibrated rules can introduce information that cannot be fully recovered, or allow eliminations that presuppose more than the introduction justified. By contrast, DNI and DNE exhibit perfect balance. Consider the schematic demonstration:

```
    A
_____¬¬I
¬¬A
_____¬¬E
    A
```

This compact proof encapsulates the fact that whatever is gained through double negation introduction can be discharged without loss by double negation elimination. It represents not a circular derivation but a guarantee that double negation does not extend the inferential space beyond the original assertion. The proof-theoretic significance of this cannot be overstated. Harmony prevents double negation from distorting the inferential landscape, while simultaneously securing the legitimacy of RAA. It is precisely because of this harmony that reductio proofs are not inferentially excessive but instead respect the internal architecture of classical logic. Thus, the benefit of harmony here is not confined to the assurance of balance between DNI and DNE; it also ensures that the inferential structure of classical logic can absorb reductio as a legitimate, non-distorting procedure. In this way, harmony demonstrates that the classical system is not merely semantically complete but proof-theoretically well-calibrated.

The companion property of normalisation reinforces these insights by showing that the use of double negation does not burden proofs with unnecessary detours. Normalisation, in proof theory, is the process of reducing derivations to their canonical forms, eliminating superfluous steps without losing validity. Within the context of DNI and DNE, normalisation demonstrates that derivations involving the introduction and subsequent elimination of double negation can always be simplified to a direct proof of the conclusion. Consider the following reduction:

```
    A
_____¬¬I
¬¬A
_____¬¬E         Normalises to:      Π
    A                                A
```

Here the detour from A to ¬¬A and back to A is reduced to a straightforward derivation of A. This collapse is not trivial but reveals an essential structural economy: the use of double negation never creates unresolvable loops or redundant derivations. Instead, it guarantees that every application of DNI followed by DNE normalises to the original assertion. This property reveals that the rules are not only harmonious but also efficient, ensuring that proofs do not accumulate unnecessary complexity. Furthermore, the claim that every instance of DNI followed by DNE normalises to the original assertion can also be demonstrated through the following sequent calculus proof:

```
           A ⇒ A                                    A ⇒ A
        ─────────── L¬                           ─────────── R¬
          A,¬A ⇒                                   ⇒ A,¬A
        ─────────── R¬                           ─────────── L¬
          A ⇒ ¬¬A                                   ¬¬A ⇒ A
        ────────────────────────────────────────────────────── Cut
                              A ⇒ A
```

where the use of cut allows us to derive the initial sequent of the proof namely, A ⇒ A. Thus, normalisation protects the proof system from inferential stagnation and demonstrates that the deployment of double negation preserves, rather than compromises, the economy of reasoning.

The proof-theoretic benefits of normalisation are manifold. First, it guarantees the subformula property in derivations involving double negation: proofs do not require the manipulation of formulae more complex than those already present. Secondly, it secures the termination of derivations, ensuring that no infinite regress arises from repeated applications of ¬¬I and ¬¬E. Thirdly, and perhaps most importantly, it underpins the process of cut-elimination, the cornerstone of Gentzenian proof theory. The elimination of cuts depends on the ability to collapse detours and reduce derivations to simpler forms. By demonstrating that double negation rules are normalisable, we confirm that they integrate seamlessly into this broader structural discipline. Far from being an obstacle, DNI and DNE actively support the consistency and clarity of the system.[16] Harmony ensures that their inferential role is balanced, while normalisation guarantees their procedural economy. Together, these properties elevate DNI and DNE from mere operational conveniences to exemplars of proof-theoretic elegance and discipline within classical logic.

In addition to these formal advantages, the harmonious and normalisable character of DNI and DNE also carries deeper conceptual significance. Proof theory is not merely concerned with whether an inference is valid, but also with whether the rules that produce it are disciplined, transparent, and resistant to excess. A rule that permits inferential overreach threatens the consistency of the system; one that cannot be normalised risks undermining its clarity. By contrast, DNI and DNE embody precisely those properties that ensure inferential discipline. They demonstrate that classical logic, while granting the full power of RAA, does so in a manner that is structurally contained. The combination of harmony and normalisation secures the result that the move from A to ¬¬A, and back from ¬¬A to A, does not inflate the inferential landscape or jeopardise the transparency of derivations. Instead, this movement exemplifies the internal stability of classical consequence. The interderivability of A and ¬¬A is not only semantically guaranteed by bivalence but also structurally underwritten by the very best proof-theoretic virtues. This convergence of semantics and proof theory provides the system with its robustness: classical logic is not only materially adequate in capturing truth but also formally adequate in maintaining disciplined proofs.

---

[16] In philosophical terms, this reinforces their legitimacy: they are not admitted merely because semantic equivalence dictates that A and ¬¬A coincide, but because they display the very structural virtues proof theory values most.

Finally, it is worth emphasising that the presence of harmony and normalisation in DNI and DNE illuminates the role of double negation as a bridge between indirect and direct reasoning. The reductio method often raises concerns that it is an inherently indirect procedure, dependent on detours through contradiction. Yet the analysis presented here shows that, when articulated through DNI and DNE, reductio operates with both balance and economy.[17] Harmony ensures that the steps taken in reductio do not exceed their justificatory basis, while normalisation ensures that the detours it generates can always be collapsed into simpler derivations. The result is that reductio, far from being an awkward appendage to classical logic, is structurally integrated into its inferential fabric. Thus, double negation plays a decisive role: it is the mechanism through which reductio becomes harmonious and normalisable. This finding not only deepens our understanding of DNI and DNE but also offers a broader insight into the architecture of classical logic. As we now move toward the conclusion of this paper, the point to stress is that double negation is not a peripheral curiosity. Through its harmonious and normalisable rules, it reveals the proof-theoretic elegance and discipline of the classical system, showing that RAA can be fully integrated into a rigorous and stable inferential framework.

## 3 Concluding Remarks

The arguments presented throughout this paper have sought to illuminate the proof-theoretic foundations of double negation introduction and elimination, situating them within the broader classical landscape. At the heart of the discussion lies the identity relation A ↔ ¬¬A, which we have shown to be indispensable both semantically and proof-theoretically. From the semantic perspective, this identity encapsulates the principle of bivalence: the truth-value of A is always identical to the truth-value of ¬¬A, guaranteeing that propositions remain stable under the operation of double negation. From the proof-theoretic perspective, the equivalence between A and ¬¬A provides the necessary foundation for reductio ad absurdum, ensuring that contradiction can be transformed into positive assertion. Without this equivalence, RAA would remain an incomplete tool, capable only of refuting negations but not of securing assertions. By contrast, with DNI and DNE functioning as the operationalisation of this identity, classical logic achieves the full scope of inferential stability. Our examination has demonstrated that DNI derives from the discharge of [¬A] to establish ¬¬A, while DNE collapses [¬¬A] back into A through reductio. These processes exemplify how RAA is embedded in the rules themselves, thereby transforming double negation from a formal curiosity into a structural cornerstone of classical logic.

---

[17] In the near future, this controversial claim may provide us with the opportunity to satisfy, at least in part, some of the expectations of intuitionistic logicians namely, by putting forward well-behaved direct classical proofs through the application of DNI and DNE.

Building on this, the paper has explored the derivability of DNI and DNE within both sequent calculus and natural deduction, showing that they originate in the structural resources of RAA. The sequent calculus proofs illustrate how both rules extend from the reflexive sequent A ⇒ A, thus preserving identity across the introduction and elimination of double negation. In natural deduction, DNI is shown to arise from assuming [A], deriving contradiction with [¬A], and discharging the latter to obtain ¬¬A; conversely, DNE is demonstrated by assuming [¬¬A], deriving contradiction with ¬A, and collapsing the negation to restore A. In each case, the rules do not operate in isolation but rather as complementary halves of a single structural process grounded in identity. Their status as natural deduction rules, ¬¬I and ¬¬E, reinforces the sense in which double negation is a proof-theoretic extension of the identity axiom, rather than an independent or accidental device. Furthermore, the derivations confirm that the validity of both DNI and DNE does not rest on semantic stipulation alone but is internally justified within the inferential machinery of proof theory. The correspondence between sequent calculus and natural deduction therefore provides a dual assurance: that double negation is both derivationally secure and conceptually indispensable. In this sense, the rules exemplify the convergence of proof-theoretic and semantic considerations that gives classical logic its distinctive character.

A final set of arguments has emphasised the structural virtues of DNI and DNE, particularly their harmony and normalisability. Harmony guarantees that introduction and elimination do not exceed or undercut one another: what DNI introduces, DNE can eliminate, and what DNE eliminates is precisely what DNI introduced. This reciprocity ensures that the rules operate within a balanced inferential economy, preventing either rule from generating content that cannot be managed by its counterpart. The schematic derivation of:

```
   A
─────¬¬I
  ¬¬A
─────¬¬E
   A
```

encapsulates this equilibrium, showing that double negation does not extend inference beyond the original assertion. Normalisation, in turn, guarantees that derivations involving DNI followed by DNE collapse into direct proofs, thus preventing detours from inflating the complexity of reasoning. The reduction of:

```
   A
─────¬¬I
  ¬¬A
─────¬¬E          to:          Π
   A                           A
```

demonstrates that the detour through double negation is always dispensable, thus preserving the economy and transparency of proofs. These virtues are not merely technical refinements but foundational guarantees of proof-theoretic discipline. They ensure that RAA, often criticised as a non-constructive or indirect method, operates within classical logic in a manner that is both rigorous and stable. The combination of harmony and normalisation thus elevates DNI and DNE to exemplars of proof-theoretic elegance: they embody the very standards of discipline, transparency, and economy that underpin Gentzen's programme of cut elimination and proof reduction.

Looking ahead, the results of this study suggest several promising directions for future research in both proof theory and the philosophy of logic. From the proof-theoretic perspective, further work could investigate whether the properties of harmony and normalisation, demonstrated here for DNI and DNE, extend to other fundamental rules of classical logic, such as those governing disjunction or even structural principles like contraction and cut. Such an extension would clarify whether the structural virtues identified in double negation are local to negation or representative of a deeper discipline in the inferential fabric of classical logic. From the philosophical perspective, the finding that DNI and DNE are harmonious and normalisable reframes RAA as a structurally disciplined rather than an excessively indirect procedure, thereby challenging the long-standing intuitionistic critique of classical logic. This in turn invites reconsideration of the boundary between classical and constructive logics, particularly regarding the legitimacy of RAA as a principle of positive assertion. A further avenue lies in exploring how these results can be integrated into bilateralist and substructural frameworks, where the force of assertion and denial is central. In such contexts, the proof-theoretic character of double negation may provide a testing ground for theories of consequence that balance semantic adequacy with inferential economy, ultimately contributing to a more unified account of the architecture of logical systems.